\documentclass{pasj00}

\begin{document}
\SetRunningHead{J.~T. Mariska et al.}{Active Region Dynamics}
\Received{2007/06/05}
\Accepted{2007/08/29}

\title{Hinode EUV Imaging Spectrometer Observations of Solar
  Active Region Dynamics}

\author{John T. \textsc{Mariska},\altaffilmark{1}
Harry P. \textsc{Warren},\altaffilmark{1}
Ignacio \textsc{Ugarte-Urra},\altaffilmark{2}
David H. \textsc{Brooks},\altaffilmark{2}
David R. \textsc{Williams},\altaffilmark{3}
\and
Hirohisa \textsc{Hara}\altaffilmark{4}}

\altaffiltext{1}{Space Science Division, Naval Research Laboratory,
  Washington DC 20375, USA} \email{mariska@nrl.navy.mil}
\altaffiltext{2}{George Mason University, 4400 University Drive, Fairfax,
  VA 22020, USA}
\altaffiltext{3}{Mullard Space Science
  Laboratory, University College London, Holmbury St. Mary,
  Dorking, Surrey, RH5 6NT, UK}
\altaffiltext{4}{National Astronomical
  Observatory, Mitaka, Tokyo 181-8588}

\KeyWords{Sun: corona --- Sun: oscillations --- Sun: UV
  radiation}

\maketitle

\begin{abstract}
The EUV Imaging Spectrometer (EIS) on the Hinode satellite is
capable of measuring emission line center positions for Gaussian
line profiles to a fraction of a spectral pixel, resulting in
relative solar Doppler-shift measurements with an accuracy of
less than a km~s$^{-1}$ for strong lines. We show an example of
the application of that capability to an active region
sit-and-stare observation in which the EIS slit is placed at one
location on the Sun and many exposures are taken while the
spacecraft tracking keeps the same solar location within the
slit. For the active region examined (NOAA 10930), we find that
significant intensity and Doppler-shift fluctuations as a
function of time are present at a number of locations. These
fluctuations appear to be similar to those observed in
high-temperature emission lines with other space-borne
spectroscopic instruments. With its increased sensitivity over
earlier spectrometers and its ability to image many emission
lines simultaneously, EIS should provide significant new
constraints on Doppler-shift oscillations in the corona.
\end{abstract}

\section{Introduction}

The EUV Imaging Spectrometer (EIS) on Hinode uses multilayer
optics to produce stigmatic spectra in two 40~\AA\ wide EUV bands
centered at 195 and 270~\AA. Emission lines present in these
bands originate in the solar transition region and corona,
permitting detailed study of the temperature, density, and
dynamical structure of these regions. The Hinode mission is
described in \citet{Kosugi2007}. An overall description of EIS is
available in \citet{Culhane2007}. \citet{Lang2006} provides more
detailed information on the EIS calibration, and
\citet{Korendyke2006} provides details of the optics and
mechanisms.

EIS can image the Sun using 1\arcsec\ and 2\arcsec\ slits and
40\arcsec\ and 266\arcsec\ slots. The latter result in
monochromatic images of the Sun in the stronger spectral lines.
Obtaining the full diagnostic capability of the instrument,
however, is only possible when data are acquired with the
1\arcsec\ or 2\arcsec\ slits. This means that images must be
constructed by moving the EIS mirror to raster an area of
interest. There is, of course, a scientific tradeoff to be made
for constructing a spectroheliogram in this manner---it can take
considerable time. For an exposure time of 30~s, a typical value
for capturing both the strongly- and weakly-emitting portions of
an active region, it can take an hour or more to image a
substantial area of the disk.

When high time cadence is important, a better approach is to use
EIS in a sit-and-stare mode. In this case, the 1\arcsec\ or
2\arcsec\ slit is placed at one location on the Sun and repeated
exposures are obtained while the Hinode spacecraft tracks the
solar rotation so that the slit always samples the same solar
plasma. These kinds of observations can be enhanced by preceding
and/or following them with EIS spectroheliograms, and are, of
course, further enhanced by using imaging data from the Hinode
X-Ray Telescope (XRT) to monitor the overall evolution of the
area being examined with the sit-and-stare observation. Details
of the capabilities of the XRT on Hinode are available in
\citet{Golub2007}.

In this paper, we present some initial results from one EIS
sit-and-stare data set. The focus here is on small fluctuations
in an active region. \citet{Imada2007} discuss additional EIS
dynamical observations during the gradual phase of an X-class
flare.

\section{Observations}

The observations discussed in this contribution cover portions of
NOAA active region 10930 and were obtained on 2006 December 14,
when it was located at approximately S05W39. EIS
spectroheliograms covering portions of the active region were
obtained beginning at 3:23 and 19:20~UT. Each spectroheliogram
consisted of 256 30~s exposures with the 1\arcsec\ slit and a
spatial step size between exposures of 1\arcsec, taking
approximately 2~hr 23~m to complete. A 120-exposure EIS
sit-and-stare observation within the region was begun at
17:00~UT. That study used 60~s exposures.

For both the spectroheliograms and the sit-and-stare
observations, nine data windows were selected on the EIS
detectors. This paper only presents results for the lines in six
of those windows. Each window in the spectroheliograms was 24
spectral pixels wide and covered a 256\arcsec\ region of the Sun
in the N/S direction. For the sit-and-stare data set, each window
was 32 spectral pixels wide, but the slit covered a
512\arcsec\ region in the N/S direction. Each EIS spectral pixel
is 22.3~m\AA\ wide. The measured FWHM at 185~\AA\ is
47~m\AA\ \citep{Culhane2007}. Table~\ref{table:lines} lists the
emission lines included in this study and their temperatures of
formation.

\begin{table}
\caption{Emission lines observed with EIS}\label{table:lines}
\begin{center}
\begin{tabular}{lll}
\hline
Ion & Wavelength (\AA) & Log $T_\mathrm{max}$ (K) \\
\hline
Fe~\textsc{viii} & 185.21 & 5.56 \\
Fe~\textsc{x} & 184.54 & 5.98 \\
Fe~\textsc{xii} & 195.12 & 6.11 \\
Fe~\textsc{xiii} & 202.04 & 6.20 \\
Fe~\textsc{xiv} & 274.20 & 6.28 \\
Fe~\textsc{xv} & 284.16 & 6.32 \\
\hline
\end{tabular}
\end{center}
\end{table}

\begin{figure}
\begin{center}
\FigureFile(2.70in,2.70in){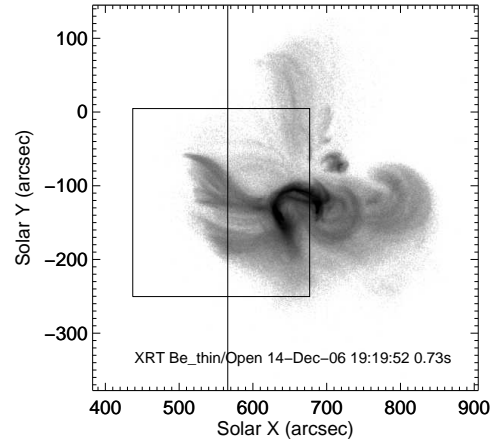}
\end{center}
  \caption{An XRT image of NOAA active region 10930 showing the
    location of the EIS spectroheliogram taken from 19:20:12 to
    21:34:55~UT, and the slit location for the EIS sit-and-stare
    observation taken from 17:00:20 to 19:04:06~UT. The
    coordinate system on the figure corresponds to the time of
    the XRT exposure and the EIS spectroheliogram and slit
    location have been rotated to that time.}
\label{fig:xrt_full}
\end{figure}

Figure~\ref{fig:xrt_full} shows an XRT image of the active region
taken at 19:19:52 UT with a box marking the region covered by the
EIS spectroheliogram that followed the sit-and-stare observation
and a vertical line showing the position of the EIS slit during
the sit-and-stare observation. The image was obtained with the
XRT Be\_thin/Open filter combination, which has peak sensitivity at
a temperature of roughly 10~MK and is two orders of magnitude
less sensitive at about 2~MK \citep{Golub2007}. Thus, the image
is mostly dominated by emission at higher temperatures than
those seen in most of the EIS emission lines used in this study.

All the data were processed using the current version of the EIS
data preparation software. This removed the detector bias and
dark current, as well as hot pixels and cosmic ray hits, and then
applied the EIS absolute calibration, yielding intensities in
ergs cm$^{-2}$ s$^{-1}$ sr$^{-1}$ \AA$^{-1}$. The software
replaces the hot pixels and those affected by cosmic ray hits
with interpolated values from adjoining pixels, but the data at
those locations were flagged as bad and the line fitting software
gave those locations no weight in the subsequent line profile
fitting process. The data on the long-wavelength detector were
also shifted in the y-direction to account for the roughly 17
pixel offset between the two detectors.

\begin{figure*}
\begin{center}
\FigureFile(6.3in,4.5in){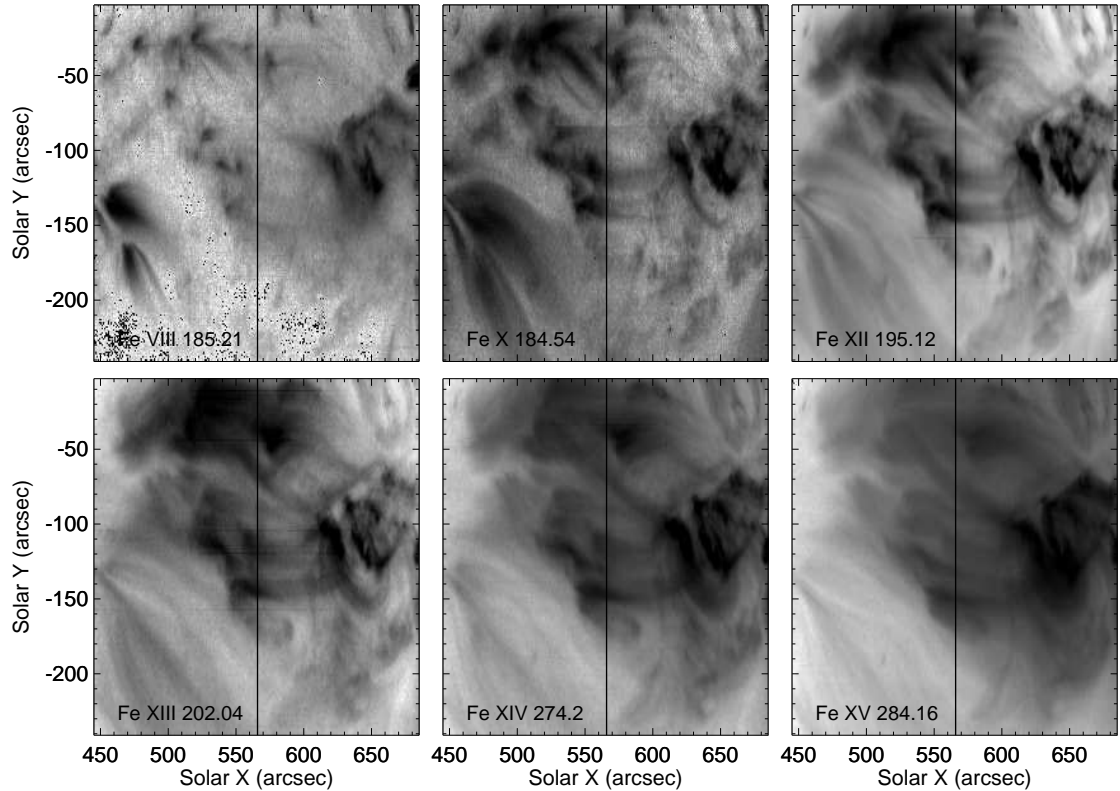}
\end{center}
  \caption{EIS spectroheliograms in six different emission lines
    obtained from 19:20:12 to 21:34:55~UT on 2006 December 14.
    The vertical line in each plot shows the position of the EIS
    1\arcsec\ slit during the sit-and-stare observation, which
    was taken from 17:00:20 to 19:04:06~UT. The coordinate system
    for the figure corresponds to the time of the middle of the
    spectroheliogram.}
\label{fig:spectroheliogram}
\end{figure*}

Figure~\ref{fig:spectroheliogram} shows the appearance of the
active region in the EIS spectroheliograms taken immediately
after the sit-and-stare study. The vertical line on each panel
shows the location of the EIS slit during the sit-and-stare
observation. As the images show, the portions of the active
region imaged by EIS had a complex structure. Examination of
movies made from XRT images of the active region shows that
considerable flare-like activity was taking place throughout the
time of the EIS observations. Data from the GOES X-ray monitors
also show significant flare-like fluctuations at the B to C
level, and an X-flare took place later in the day. The bright,
compact loops in the XRT image are the result of a small flare
that took place before the EIS slit reached those positions. By
the time the EIS slit reached those positions, the strong
emission had decayed and the loop morphology is less evident.
(EIS spectroheliograms are constructed by scanning the slit from
west to east on the Sun.)

\section{Intensity and Doppler-shift fluctuations}

\begin{figure*}
\begin{center}
\FigureFile(6.3in,4.5in){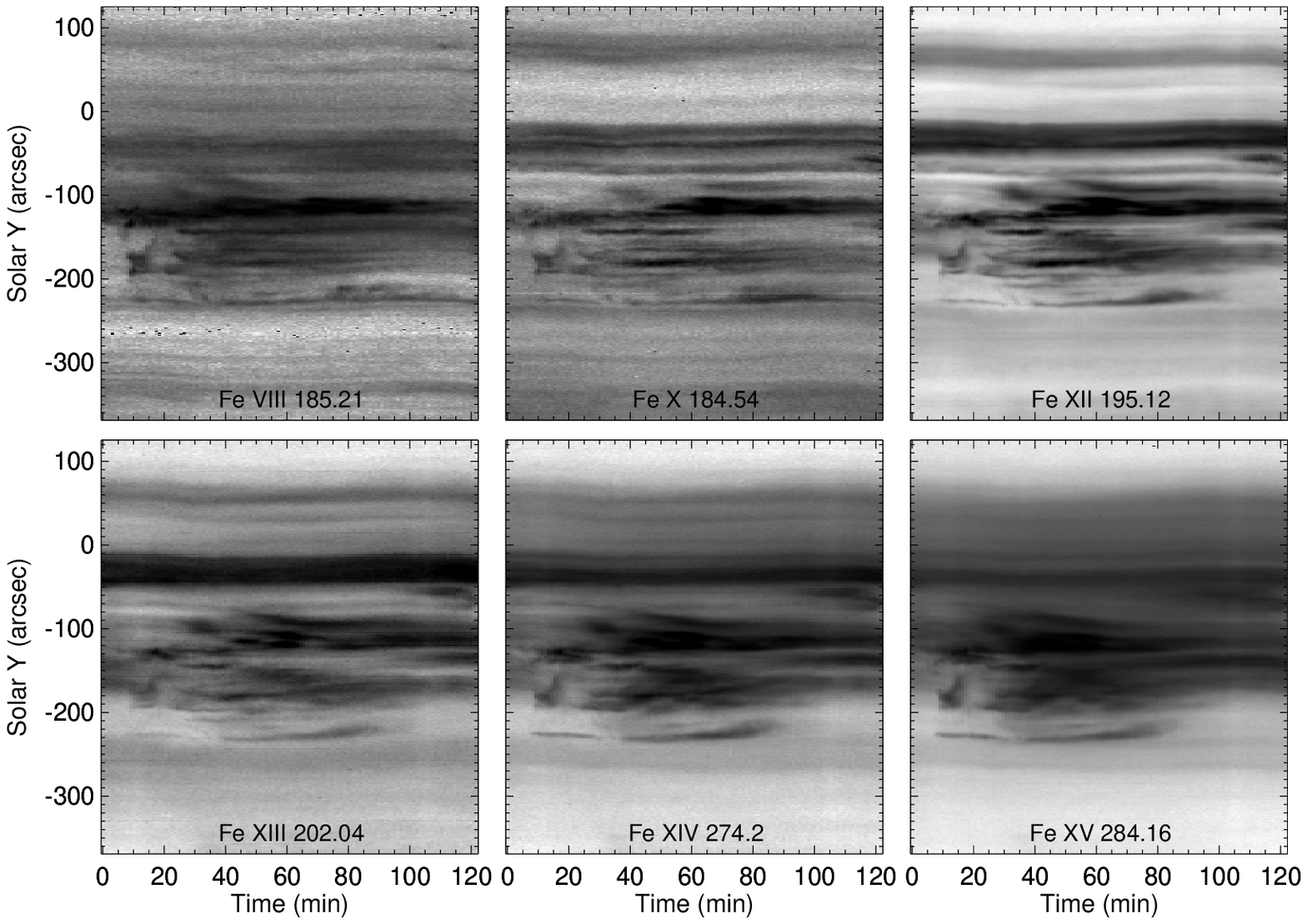}
\end{center}
  \caption{EIS sit-and-stare intensity data in six different
    emission lines obtained beginning at 17:00:20~UT on 2006
    December 14. The horizontal axis values are the start times
    of the exposures.} 
\label{fig:sns-image}
\end{figure*}

For the sit-and-stare observation, the emission lines in the six
wavelength bands shown in Figure~\ref{fig:spectroheliogram} were
all fitted with single Gaussian line profiles. Thus, for each
spectral line at each location along the EIS slit, we have
measured values of the total line intensity, the location of the
line center, and the line width. Figure~\ref{fig:sns-image} shows
a greyscale representation of the resulting total intensity in
each spectral line as a function of time for all 512 positions
along the slit. A number of locations show obvious evidence for
rapid time variations in the total intensity. Comparison with the
EIS spectroheliograms in Figure~\ref{fig:spectroheliogram} shows
that these locations appear to correspond generally to areas near
the footpoints of loops. Examination of movies made from the XRT
images and those from the EUV Imaging Telescope (EIT) on the Solar
and Heliospheric Observatory (SOHO) suggest that these are due to
intermittent heating events in some of the loops crossed by the
EIS slit.

Extracting Doppler-shift information is a somewhat complex
undertaking. The EIS slit is not perfectly vertical on the camera
CCDs, so a correction for the slit tilt must be made. In
addition, over the course of each orbit, EIS exhibits a periodic
variation in the measured line center position for all spectral
lines observed in both detectors. The peak-to-peak variation is
about 2 spectral pixels---a significant Doppler signal at the
wavelengths covered by EIS. The cause of this variation is still
under investigation, but appears to be due to temperature changes
within the instrument.

The slit tilt does not appear to change with time and results in
a difference in the location of the rest wavelength for an
emission line of approximately 0.37 spectral pixels from the
bottom to the top of the 512\arcsec\ slit length used in the
sit-and-stare observations. While this is small, it represents a
velocity change of about 13~km s$^{-1}$ at the Fe~\textsc{xii}
195.12~\AA\ emission line and thus is important for detailed
dynamical studies. The slit tilt was removed from the
sit-and-stare data set by shifting the fitted line center
position in each spectral line by an amount determined using an
analysis of Doppler-shift data taken early in the mission.

To a first approximation, the orbital variation is the same at
each location on the two EIS detectors. Thus, it is possible to
use the data in a high-signal-to-noise channel during the
sit-and-stare observation to determine the orbital variations for
all the wavelengths. To do this we use the top 112 rows of fitted
line center positions (corrected for slit tilt) in the
Fe~\textsc{xii} 195.12~\AA\ emission line data to determine an
average orbital variation. This orbital variation is then
subtracted from the line center positions measured in all the
wavelength windows. This particular region was selected because
examination of the XRT and EIT data indicates that no significant
dynamical activity is present at these locations. Note that EIS
has no absolute wavelength reference. Once the orbital trend has
been removed using the Fe~\textsc{xii} 195.12~\AA\ emission line
data, the average value of the line center position in the top
112 rows in each of the six emission lines used in this study was
used as the rest wavelength for that emission line.

\begin{figure*}
\begin{center}
\FigureFile(6.3in,4.5in){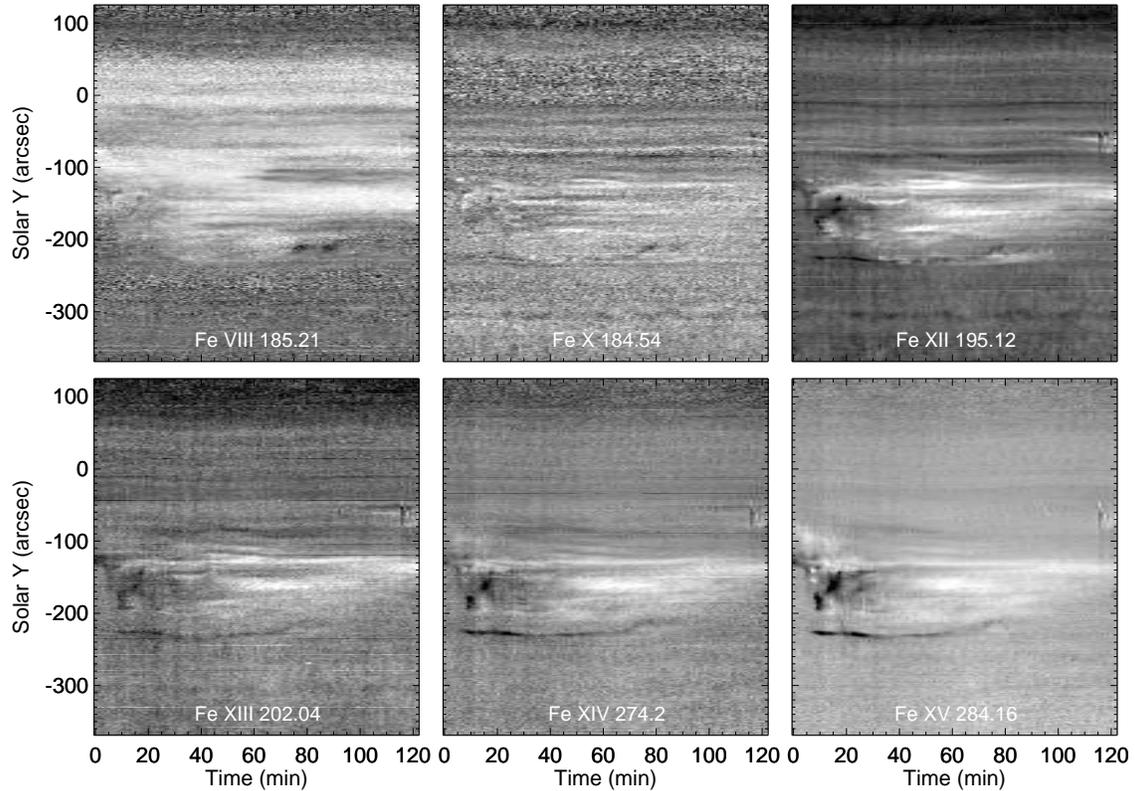}
\end{center}
\caption{EIS sit-and-stare Doppler-shift data in six different
  emission lines obtained beginning at 17:00:20~UT on 2006
  December 14. The horizontal axis values are the start times of
  the exposures. Darker values represent smaller numbers
  (blueshifts). The zero point for the Doppler shifts is
  arbitrary. The range of Doppler shifts is 374 km s$^{-1}$ for
  Fe~\textsc{viii}, 88 km s$^{-1}$ for Fe~\textsc{x}, 28 km
  s$^{-1}$ for Fe~\textsc{xii}, 25 km s$^{-1}$ for
  Fe~\textsc{xiii}, 42 km s$^{-1}$ for Fe~\textsc{xiv}, and 266
  km s$^{-1}$ for Fe~\textsc{xv}.}
\label{fig:sns-Doppler}
\end{figure*}

Figure~\ref{fig:sns-Doppler} shows a greyscale representation of
the fitted line center positions with the orbital variations
removed as a function of time for all 512 positions along the
slit for the six lines included in this study. In these images
darker shades represent smaller values for the Doppler shift
(blueshifts). If one excludes data points that are obvious
outliers because of poor Gaussian fits, the range of
Doppler-shift velocities seen in the panels is on the order of 20
to 100 km~s$^{-1}$.

Examination of the sit-and-stare images shown in
Figures~\ref{fig:sns-image} and \ref{fig:sns-Doppler} shows that
all the significant features shown along the slit location in the
the spectroheliograms shown in Figure~\ref{fig:spectroheliogram}
are present in the sit-and-stare data. Near the top of this
$Y\!$-range (roughly $-10$\arcsec\ to $-40$\arcsec, there are
nearly constant intensity bands in all the emission lines.
Examination of the spectroheliograms and especially of a movie
made from the EIT 195~\AA\ images shows that these are quiescent
loops interconnecting portions of the active region. The
Doppler-shift sit-and-stare data at these locations show little
or no time-dependent fluctuation in any of the emission lines
observed with EIS.

In contrast, the $Y\!$-range from $-100$\arcsec\ to
$-230$\arcsec\ shows considerable time-dependent behavior in both
the line intensity and the Doppler shift. While the intensity
variations are visible in all the emission lines, there is a
clear tendency for the Doppler shifts to become more pronounced
as the temperature of formation of the emission line increases.

\begin{figure}
\begin{center}
\FigureFile(3.15in,3.15in){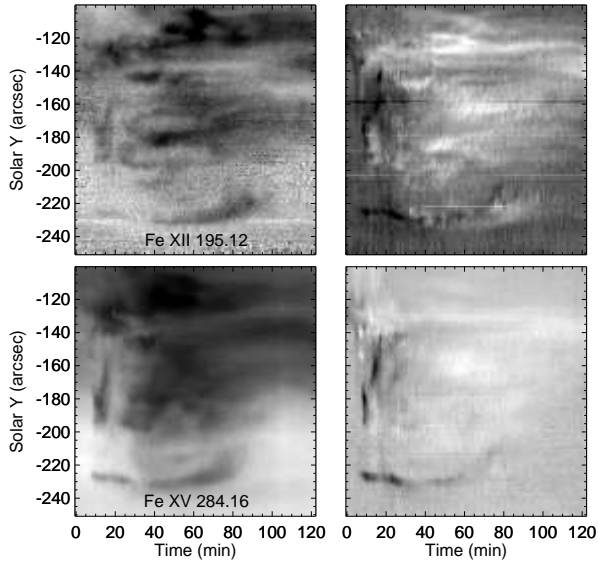}
\end{center}
\caption{EIS sit-and-stare intensity (left panels) and
  Doppler-shift (right panels) data for Fe~\textsc{xii} and
  Fe~\textsc{xv} for a portion of slit range covered by
  Figures~\ref{fig:sns-image} and \ref{fig:sns-Doppler}. For the
  Doppler shift data, darker values represent smaller numbers
  (blueshifts). The zero point for the Doppler shifts is
  arbitrary. The range of Doppler shifts is 87 km s$^{-1}$ for
  Fe~\textsc{xii} and 96 km s$^{-1}$ for Fe~\textsc{xv}.}
\label{fig:sns-expanded}
\end{figure}

Figure~\ref{fig:sns-expanded} shows greyscale images of the
Fe~\textsc{xii} and Fe~\textsc{xv} intensity and Doppler-shift
data along the portion of slit where considerable activity is
taking place. Some features are clearly impulsive, such as the
upflow at $-180$\arcsec\ and 10~min and a second at
$-160$\arcsec\ and 15~min seen in the Fe~\textsc{xii} and
Fe~\textsc{xv} lines. More gradual changes in the Doppler shifts
are also evident. For example, in both the Fe~\textsc{xii} and
Fe~\textsc{xv} Doppler images in the vicinity of $-160$ and
60~min, the redshift velocities appear to be changing gradually
over 20-min and longer timescales.

\begin{figure}
\begin{center}
\FigureFile(3.15in,2.88in){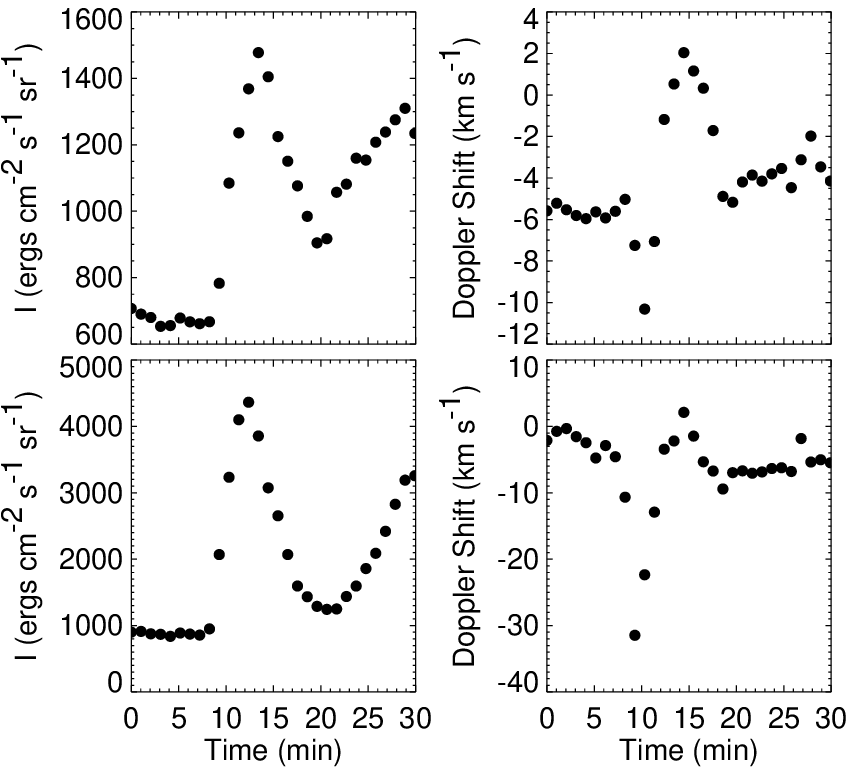}
\end{center}
\caption{EIS sit-and-stare intensity and Doppler-shift data for
  Fe~\textsc{xii} (top row) and Fe~\textsc{xv} (bottom row) for a
  solar $Y\!$-position near $-185$\arcsec. Negative values for
  the Doppler shift represent blueshifted emission. The
  zero-velocity point has been arbitrarily set to the average
  value for all 120 measurements at this $Y\!$-position.}
\label{fig:sns-185}
\end{figure}

Figure~\ref{fig:sns-185} shows intensity and Doppler-shift data
in the Fe~\textsc{xii} and Fe~\textsc{xv} emission lines as a
function of time at a solar $Y\!$-position of approximately
$-185$\arcsec. The graphs quantify the impression given in
Figure~\ref{fig:sns-expanded}---an impulsive event of some kind
has taken place causing a relative blueshift followed by a
relative redshift. In the Doppler-shift plots and to a lesser
degree in the intensity plots, the events appear to be
superimposed on a more smoothly varying background.

The Doppler shifts in both the Fe~\textsc{xii} and Fe~\textsc{xv}
emission lines appear to show roughly one complete oscillation.
In the intensity plots, the full oscillation is less obvious,
possibly because the signal is superimposed on the more gradual
trend present in the data. Measuring the time between the maximum
blueshift and the maximum redshift in each line and assuming that
it corresponds $1/2$ the oscillation period, results in
oscillation periods of 8.2~min for the Fe~\textsc{xii} emission
line and 10.2~min for the Fe~\textsc{xv} emission line. Given the
one-minute cadence of the sit-and-stare observation, these
numbers should be considered roughly equal.

Measuring the time between the intensity peak and the following
local minimum and assuming that it corresponds to $1/2$ the
oscillation period results in oscillation periods of 14.4 and
16.4~min for the Fe~\textsc{xii} and Fe~\textsc{xv} emission
lines, respectively. Thus, the intensity fluctuations do not have
the same period as the Doppler-shift fluctuations. Note also that
the peak in the intensity oscillation appears to be later than
the maximum blueshift.

The characteristics of the intensity and Doppler-shift
oscillations show in Figure~\ref{fig:sns-185} are very similar to
those observed for example by \citet{Wang2003} and
\citet{Wang2005} using the SUMER spectrometer on SOHO.
\citet{Wang2005} reported average oscillation periods of $17.9
\pm 5.5$~min with a range from 8.6 to 23.3~min, comparable to the
oscillation shown in the figure. They reported a wide range of
Doppler shift amplitudes---ranging from 12 to 353 km s$^{-1}$,
with an average value of $62 \pm 57$ km s$^{-1}$. We find
amplitudes of roughly 6.5 and 25.8 km s$^{-1}$ for the
Fe~\textsc{xii} and Fe~\textsc{xv} emission lines, respectively,
within the range seen in the SUMER observations.

\begin{figure}
\begin{center}
\FigureFile(2.8in,3.2in){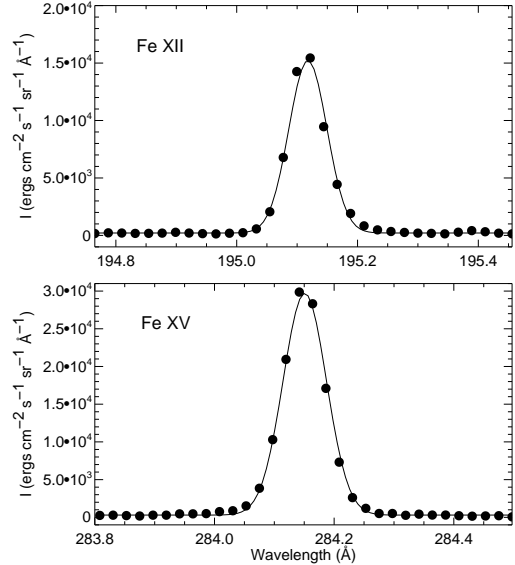}
\end{center}
\caption{Sample Fe~\textsc{xii} 195~\AA\ and Fe~\textsc{xv}
  284~\AA\ spectral window data obtained at a solar
  $Y\!$-position of $-185$\arcsec\ at 17:14:48~UT, near the peak
  in the Doppler shift plots shown in Figure~\ref{fig:sns-185}.
  The lines show the best fit Gaussian to each data set.}
\label{fig:sns-185-profile}
\end{figure}

Note that for the most part the Doppler shift velocities are
quite small, and one might ask whether EIS is capable of
measuring such small velocity changes.
Figure~\ref{fig:sns-185-profile} shows sample Fe~\textsc{xii} and
Fe~\textsc{xv} spectral data taken near the peak in the
Doppler-shift plots along with the best-fit Gaussians. The
one-$\sigma$ uncertainties on the fits are 0.44 and 0.63 km
s$^{-1}$ for the Fe~\textsc{xii} and Fe~\textsc{xv} data,
respectively. This level of accuracy is achievable with EIS
whenever the signal level is high and a Gaussian represents a
good functional form for the line profile.

\section{Conclusions}

The sit-and-stare observations presented in this paper and other
similar EIS active region observations clearly show an abundance
of dynamical phenomena in active regions, much of it periodic.
Cotemporal XRT imaging data often show that these phenomena are
the dynamical manifestation of disturbances that can be seen
following field lines that interconnect portions of the active
region. Thus, careful analysis of combined EIS and XRT data sets
should provide an improved picture of how active regions evolve.
Ultimately, the addition of chromospheric images and vector
magnetograms from the Solar Optical Telescope on Hinode should
make it possible to begin to connect fully changes in the corona
with the evolution of the underlying magnetic field.

The particular example shown in Figure~\ref{fig:sns-185} is
similar to oscillatory phenomena observed in high-temperature
emission lines with the SUMER spectrometer on SOHO. The intensity
signal shown near $-185$\arcsec\ in Figure~\ref{fig:sns-expanded}
does begin with a fairly low intensity, and one might speculate
that higher temperature plasma in emitting at that time at that
location. The data set we have analyzed for this study does not
include the hotter line of Fe~\textsc{xvii} at
204.65~\AA\ available in the EIS bandpass, and, while we do have
data for the Ca~\textsc{xvii} emission line at 192.82~\AA, the
line suffers from a blend with a cooler line of Fe~\textsc{xi},
which can only be removed with information from other
Fe~\textsc{xi} lines. Thus, a full comparison with SUMER results
must await the analysis of additional data.

Many of these events are thought to be standing slow-mode waves,
the detailed study of which should lead to a better understanding
of heating in coronal loops. Because EIS is a more sensitive
instrument than SUMER, shorter exposure times are possible, which
should allow us to expand the range of periods that can be
investigated. Moreover, the EIS wavelength bands contain lines
whose ratios are sensitive to coronal electron densities, leading
to better constraints on the characteristics of the oscillations.

Extracting subtle dynamical information from EIS data does,
however, require both carefully planned observations and an
analysis that pays attention to all the possible instrumental
effects. Many of the instrumental effects are just beginning to
be understood, so it may be some time before it is possible to
fully realize the potential EIS offers for more completely
characterizing solar coronal dynamics.

Hinode is a Japanese mission developed and launched by ISAS/JAXA,
with NAOJ as domestic partner and NASA and STFC (UK) as
international partners. It is operated by these agencies in
cooperation with ESA and NSC (Norway). We are grateful to the
entire Hinode team for all their efforts in the design and
operation of the mission. JTM, HPW, IUU, and DHB acknowledge
support from the NASA Hinode program.

\end{document}